\documentclass[twocolumn]{article}
\usepackage[utf8]{inputenc}

\usepackage{geometry}
\usepackage{soulutf8}
\usepackage{caption}
\usepackage{subcaption}
\usepackage{booktabs}
\usepackage{adjustbox}
\usepackage{multirow}
\usepackage[table]{xcolor}
\usepackage{floatrow}
\usepackage{amsmath}
\usepackage{amssymb}
\usepackage{bm}
\usepackage{nicefrac}
\usepackage{mathtools}
\usepackage[separate-uncertainty=true,parse-numbers=true]{siunitx}
\usepackage{algorithm}
\usepackage{algpseudocode}
\usepackage[colorlinks=true, linkcolor=blue, citecolor=blue, urlcolor=blue]{hyperref}
\usepackage[capitalise,nameinlink,noabbrev]{cleveref}
\usepackage[nocompress]{cite}
\usepackage{authblk}
\usepackage{pgfplots}
\usepackage{tikz}
\pgfplotsset{compat = newest}
\usetikzlibrary{quotes,angles,calc,arrows,patterns,quotes,external}
\usepgfplotslibrary{groupplots}

\renewcommand{\vec}{\boldsymbol}
\newcommand{\mat}{\mathbf}
\newcommand{\trans}{^{\mathrm{T}}}
\newcommand{\diag}{{\rm diag}}
\newcommand{\inv}{^{-1}}

\newcommand{\ex}[1]{\mathrm{E}\left[ #1 \right]} 
\newcommand{\var}[1]{\mathrm{Var}\left[ #1 \right]}

\title{Fabrication uncertainty guided design optimization of a photonic crystal cavity by
  using Gaussian processes}

\author[a]{Matthias Plock}
\author[a]{Felix Binkowski}
\author[a,b]{Lin Zschiedrich}
\author[a,b]{Philipp-Immanuel Schneider}
\author[a,b,$\dagger$]{Sven Burger}
\affil[a]{Zuse Institute Berlin, Takustraße 7, 14195 Berlin, Germany}
\affil[b]{JCMwave GmbH, Bolivarallee 22, 14050 Berlin, Germany}
\affil[$\dagger$]{Corresponding author: burger@zib.de}
\date{March 5, 2024}

\begin{document}

\maketitle

\begin{abstract}
  We present a fabrication uncertainty aware and robust design optimization approach that
  can be used to obtain robust design estimates for nonlinear, nonconvex, and expensive
  model functions. It is founded on Gaussian processes and a Monte Carlo sampling
  procedure, and assumes knowledge about the uncertainties associated with a manufacturing
  process. The approach itself is iterative. First, a large parameter domain is sampled in
  a coarse fashion. This coarse sampling is used primarily to determine smaller candidate
  regions to investigate in a second, more refined sampling pass. This finer step is used
  to obtain an estimate of the expected performance of the found design parameter under
  the assumed manufacturing uncertainties. We apply the presented approach to the robust
  optimization of the Purcell enhancement of a photonic crystal nanobeam cavity. We obtain
  a predicted robust Purcell enhancement of $\overline{F}_{\mathrm{P}} \approx 3.6$. For
  comparison we also perform an optimization without robustness. We find that an unrobust
  optimum of $F_{\mathrm{P}} \approx 256.5$ dwindles to only $\overline{F}_{\mathrm{P}}
  \approx 0.7$ when fabrication uncertainties are taken into account. We thus demonstrate
  that the presented approach is able to find designs of significantly higher performance
  than those obtained with conventional optimization.
\end{abstract}

\section{Introduction}
\label{sec:intro}

The development of new nanophotonic devices is a challenging task that is fueled
especially by advances in computing power and powerful algorithms -- both, for
optimizations~\cite{wang2021intelligent,jensen2011topology,topopt_tut,mao2021inverse,plock2022bayesian}
and for solving Maxwell's
equations~\cite{Pomplum_NanoopticFEM_2007,taflove2005computational,Iqbal21}. Nowadays,
many innovations in the field stem from inverse design
approaches~\cite{molesky2018inverse,mao2021inverse}. Here, parameterized computer models,
typically numerically discretized models of the envisioned devices are created, e.g., by
using the finite element method, and the parameters of these models are varied
systematically such that a design is identified that results in a device with optimal
figures of merit. This simulation and optimization approach allows to explore a large
number of possible device configurations and is much more economical than pure laboratory
development.

The modeled nanophotonic devices coming out of these processes often have optimized
figures of merit, e.g., maximized quality factor of an optical resonator, with record
breaking performance values~\cite{asano2017photonic}. When it is time for the
manufacturing step of the device, however, the performance values are usually not realized
to the predicted extent. This happens even in the cases where the computer model
faithfully captures the physical reality and where numerical discretization errors are
controlled to sufficiently low levels. Instead, the reason for the deviation is frequently
cited as lacking control over the process windows of the manufacturing
method~\cite{minkov2014automated,deotare2009high,hagino2009effects}. A possible
explanation for this is that the very good theoretical results found using the computer
model are due to, e.g., fragile resonance effects~\cite{Oskooi:12} where the results are
part of a narrow ridge in the parameter space, that vanish if some of the model parameters
are varied only slightly.

In some instances, to deal with the issue, researchers have employed a mass production
approach in which a large number of devices were manufactured and subsequently quantified
-- a step after which only viable samples were retained~\cite{bracher2015fabrication}.
This approach leaves much to be desired in an industrial context, where it is favorable to
go from computer model to physical device and have consistent and coherent results at each
design step. Rejecting a large number of samples is simply not economical. This
requirement adds a new layer of complexity to the design process.

Finding optima that preserve their quality in the face of model parameter uncertainties is
the task of robust optimization~\cite{park2006robust,BEYER20073190,ElsawyLPR}. Here, a
large focus is placed on convex and linear
problems~\cite{ben1998robust,gabrel2014recent,gorissen2015practical}, in which one often
attempts to optimize for the worst case. However, linear convex optimization methods are
typically not useful for the robust optimization of computer model based designs, since
these models can generally not assumed to be linear or convex. For the robust optimization
of these more general problems, e.g., non-linear and/or non-convex, heuristic approaches
are often
proposed~\cite{mutapcic2009robust,robust_design_polynomial_chaos,Oskooi:12,gostimirovic2023improving,pozzi2023robust,Wiecha:21}.
These, e.g., aim to improve robustness by considering what process input parameters would
\emph{actually} result in the desired outcome, rather than assuming that the process
produces outputs that are true to the inputs~\cite{gostimirovic2023improving}.

\begin{figure}[h]
  \includegraphics{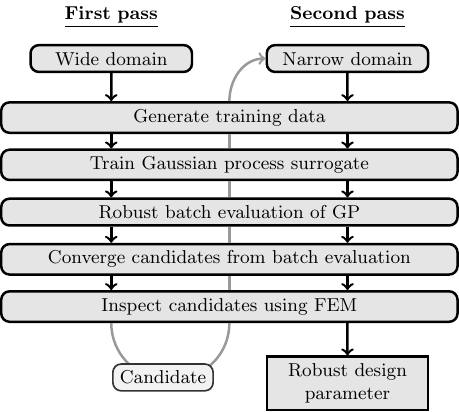}
  \caption{\label{fig:schematic} A schematic of the optimization procedure. First, the
    complete, \emph{wide domain} is considered. Training data is generated and used to
    train a Gaussian process surrogate. Then, robust estimates of the design performance
    on the complete domain are obtained by calculating Monte Carlo integrals on the
    surrogate model. The best results are considered as candidates, which are inspected
    more closely using the FEM model. Here, we select the best result and use its
    parameters to narrow the complete domain to a much more focused domain. We use this
    \emph{narrow domain} for a second pass through the pipeline to obtain robust design
    parameters, for a design which is insensitive to fabrication uncertainties. The
    complete procedure is described in more detail in \cref{sec:procedure}.}
\end{figure}

In this article, we present an iterative optimization procedure that can be used to find
robust optima in a parameterized computer model, illustrated in \cref{fig:schematic}.
The approach is founded on well understood Monte Carlo
sampling~\cite{robert1999monte,murphy2012machine} and assumes knowledge about the type and
extent of the deviations contained in the manufacturing processes parameters, such that we
can construct a probability distribution to draw samples from, e.g., a parameter $p_{i}$
scatters according to a normal distribution with some mean $\mu_{i}$ and standard
deviation $\sigma_{i}$, i.e., $p_{i} \sim \mathcal{N}(\mu_{i}, \sigma_{i})$. As such, the
presented approach leans towards branches of robust optimization called stochastic
programming~\cite{kall1994stochastic,birge2011introduction} and scenario
analysis~\cite{kall1994stochastic}. A Gaussian process machine learning surrogate
model~\cite{williams2006gaussian} of the parameterized computer model is trained and used
for the Monte Carlo sampling in lieu of the expensive model. This helps to effectively
manage the computational burden normally imposed by Monte Carlo sampling procedures. An
analysis of the percentiles of the results sampled from the Gaussian process is used to
give a prediction of the results that can actually be expected from the device when
manufactured in reality. We opt to optimize for the median, however, worst case
optimizations with very small percentiles are also thinkable.

The approach presented in this article represents the background and the current state of
development of a methodology that we have previously applied to create robust designs: In
\cite{bopp2022sawfish} a novel cavity design was first optimized using a Bayesian
optimizer~\cite{garnett_bayesoptbook_2023,schneider2019benchmarking}. The robustness of
the parameter identified by the optimizer has then been assessed with a Monte Carlo
sampling approach that was accelerated using Gaussian process surrogate models. In
\cite{rickert2023high}, a series of parameter candidates were determined using a Bayesian
optimizer. A Gaussian process surrogate model was then trained in a region around each
candidate. These Gaussian processes were used to \emph{find} the most robust candidate by
optimization. The objective function computed robust results by incorporating a large
number of Monte Carlo samples that were drawn from a probability distribution informed by
the assumed uncertainties of the manufacturing process. We found that pretraining the
Gaussian process completely is more efficient than iteratively training it. It allows us
to create training data in parallel in a batch process, and also cover the computational
domain effectively. Other authors have published articles on robust optimization using
Gaussian processes (under the term 'Kriging') and Monte Carlo
sampling~\cite{martin2006monte,lee2006global}, where differences can be found, e.g., in
the used optimization scheme (both publications employed a simulated annealing approach
that additionally helped to parameterize the sample size of the Monte Carlo sampling). In
contrast to these works, we show that our approach is able to perform well on a very
uninformed domain and does not introduce a surrogate model for derived quantities, such as
a standard deviation that is to be minimized.

A more recent work \cite{Elsawy:21} incorporates Gaussian process surrogate models of an
expensive model function together with Monte Carlo methods to perform robust design
optimization. Here, a multi objective optimization is performed by applying the Expected
\emph{Hypervolume} Improvement acquisition function in a Bayesian optimization loop. This
way both the mean as well as the standard deviation of a Monte Carlo integral are
minimized at the same time.

This article extends these approaches by also considering the robustness of the parameters
investigated in the initial candidate identification step. In
\cite{bopp2022sawfish,rickert2023high}, this was done by a Bayesian optimizer, using the
expensive forward model function, where only point values were calculated. In
\cite{bopp2022sawfish}, the robustness was only assessed and not optimized, and, in
\cite{rickert2023high}, the robustness was not optimized across the complete parameter
space.

We apply the method to the robust optimization of a photonic crystal nanobeam cavity based
on cylindrical holes in an optical waveguide~\cite{joannopoulos1997photonic,Deotare2012},
where the figure of merit we aim to maximize in a robust fashion is the Purcell
enhancement, $F_{\mathrm{P}}$~\cite{Purcell_1946}. Compared to the quality factor,
$F_{\mathrm{P}}$ varies smoother across the parameter space (since it depends on both,
quality factor and mode volume~\cite{Purcell_1946,joannopoulos1997photonic}). The Purcell
enhancement is computed by using scattering simulations with a dipole emitter modeling a
point source, which is placed in the cavity and emits at an experimentally relevant
wavelength.

The article is organized as follows: \cref{sec:theory} provides details on the employed
mathematical tools. In \cref{sec:procedure}, the optimization procedure is described in
detail. In \cref{sec:example}, the model of a photonic crystal nanobeam cavity is
introduced. In \cref{sec:results}, we apply the procedure for the optimization of the
model of the photonic cystal nanobeam cavity, and also compare the obained results to a
result found without restricting to a robust solution. Finally, \cref{sec:conclusion}
summarizes the findings.

\section{Theory}
\label{sec:theory}

The presented approach relies on Gaussian processes and Monte Carlo sampling techniques.
The theoretical foundation for these methods is reviewed in the following.

\subsection{Gaussian process regression}
\label{sec:gpr}

Gaussian processes (GPs)~\cite{williams2006gaussian} are stochastic machine learning
surrogate models that can be fit to numerical training data, i.e., inputs in the form of a
set of $M$ positions $\vec{P} = \left[ \vec{p}_{1}, \dots, \vec{p}_{M} \right]\trans$ in
some $N$-dimensional parameter space $\mathcal{X} \subset \mathbb{R}^{N}$, with each
$\vec{p}_{m} \in \mathcal{X}$, and $M$ associated numerical values $\tilde{\vec{Y}} =
  \left[ \tilde{y}_{1}, \dots \tilde{y}_{M} \right]\trans$ where each $\tilde{y}_{m} \in
  \tilde{\mathcal{Y}} \subset \mathbb{R}$. After the GP has been trained on
$\tilde{\vec{Y}}$ it can then be used as an approximator or predictor of the training data
that can be evaluated at each point in $\mathcal{X}$. The training data is often generated
by evaluating some (often expensive) parameterized process, e.g., a function $f:
  \mathcal{X} \to \tilde{\mathcal{Y}}$. The process is generally considered as a black box,
and only few assumptions are made with respect to the data generated by
it~\cite{garnett_bayesoptbook_2023}. An important one is that in order to achieve a good
fit the data should appear as if it could in principle have been drawn from \emph{some}
GP~\cite{garnett_bayesoptbook_2023}. Otherwise a GP would not be able to model it
accurately. One aspect of this is that we require the data set $\tilde{\vec{Y}}$ to
(roughly) follow a normal distribution. A huge benefit of a GP is that it is typically
much cheaper to evaluate than the process used to create the training data.

Mathematically, a GP is an infinite dimensional extension of a (finite dimensional)
multivariate normal distribution~\cite{schervish2014probability}. They are defined on the
continuous domain $\mathcal{X}$. While multivariate normal distributions are fully
determined by a mean vector $\vec{\mu}$ and a positive (semi-)definite covariance matrix
$\mat{\Sigma}$~\cite{schervish2014probability}, GPs are instead fully specified by a mean
function $\mu: \mathcal{X} \to \mathbb{R}$ and a covariance kernel function $k:
  \mathcal{X} \times \mathcal{X} \to \mathbb{R}$~\cite{williams2006gaussian}. In order to be
used as an approximation method it however still needs to be trained on the training data.
Common choices for these functions are a constant mean function $\mu(\vec{p}) = \mu_{0}$
and a Mat\'{e}rn $\nicefrac{5}{2}$ kernel function~\cite{brochu2010tutorial}, i.e.,
\begin{gather}
  \label{eq:matern}
  k(\vec{p}, \vec{p}^{\prime}) = \sigma_{0} \left( 1 + \sqrt{5}r+\frac{5}{3}r^2
  \right) \times \exp \left( -\sqrt{5}r \right) \,, \\
  \text{with} \quad r = \sqrt{\sum_{i=1}^{N} \left( \frac{p_{i} -
      p_{i}^{\prime}}{l_i} \right)^2} \,. \nonumber
\end{gather}
These choices are also applied throughout this article. The quantity $r$ describes
normalized distances between parameters $\vec{p}$ and $\vec{p}^{\prime}$. Like many other
machine learning methods GPs are also controlled by means of
hyperparameters~\cite{montavon2018methods,rasmussen2003gaussian} (here prior mean value
$\mu_{0}$, prior standard deviation $\sigma_{0}$ and length scales $l_{i}$ for each
dimension of the training data $\tilde{\vec{Y}}$). However, in contrast to, e.g., deep
neural networks~\cite{rasmussen2003gaussian}, suitable values for the hyperparameters can
be efficiently determined from the training data. The hyperparameters are chosen to
maximize the likelihood of the training data~\cite{garcia2018shape}.

The predictions of a GP are made in terms of a normal distribution for each point
$\vec{p}_{\ast} \in \mathcal{X}$, i.e.,
\begin{gather}
  \label{eq:predicted_normal}
  \hat{f}(\vec{p}_{\ast}) \sim \mathcal{N}(\overline{y}(\vec{p}_{\ast}),
  \sigma^2(\vec{p}_{\ast})) \,.
\end{gather}
The hat notation is used to indicate a random variable. After the GP has been trained on
training data, the predicted mean value $\overline{y}(\vec{p}_{\ast})$ approximates the
training data and the predicted variance value $\sigma^{2}(\vec{p}_{\ast})$ describes how
much the training data is expected to scatter around the predicted mean, given some choice
of hyperparameters. For the predicted mean and variance values, we have
\begin{align}
  \label{eq:gp_pred_mean}
  \overline{y}(\vec{p}_{\ast}) & = \mu_0 + \vec{k}\trans(\vec{p}_{\ast}) \mat{K}\inv \left( \tilde{\vec{Y}} - \mu_0 \cdot \mat{1} \right) \,, \\
  \label{eq:gp_pred_var}
  \sigma^{2}(\vec{p}_{\ast})   & = \sigma_{0}^{2} - \vec{k}\trans(\vec{p}_{\ast}) \mat{K}\inv \vec{k}(\vec{p}_{\ast}) \,,
\end{align}
with $\vec{k}(\vec{p}_{\ast}) = \left[ k(\vec{p}_{\ast}, \vec{p}_{1}), \dots,
    k(\vec{p}_{\ast}, \vec{p}_{M}) \right]\trans$, $(\mat{K})_{ij} = k(\vec{p}_{i},
  \vec{p}_{j})$, and identity matrix $\mat{1}$.

GPs are frequently applied in the context of Bayesian optimization, which is discussed
briefly in \cref{sec:bo}.

\subsubsection{Transformation based Gaussian processes for bounded data}
\label{sec:bounded_gpr}

Using data $\vec{Y}$ to train a GP that is clearly bounded to a specific subset of
$\mathbb{R}$, e.g., $\vec{Y} = \left[ y_{1}, \dots, y_{M} \right]$ with $y_{m} \in
  \mathcal{Y} = \mathbb{R}^{+}$, can lead to problems associated with the predictions of the
GP~\cite{snelson2003warped,lazaro2012bayesian}. Since the predictions of a GP are by
definition unbounded~\cite{garnett_bayesoptbook_2023,rasmussen2003gaussian}, they can
sometimes exceed the support of the training data, i.e., the predicted variance region or
even the predicted mean of the GP are assuming values not in $\mathcal{Y}$. This poses the
question of how useful these predictions are. This problematic behavior is exacerbated if,
e.g., a large amount of the training data is situated close to one side of $\mathcal{Y}$.

A possible way of dealing with this issue is by transforming the training data $\vec{Y}$
from its bounded domain $\mathcal{Y}$ to an unbounded co-domain $\tilde{\mathcal{Y}}$ with
a bijective transformation function $g: \mathcal{Y} \to \tilde{\mathcal{Y}}$. The
unbounded training data $\tilde{\vec{Y}} = \left[ \tilde{y}_{1}, \dots, \tilde{y}_{M}
    \right]\trans = \left[ g(y_{1}), \dots, g(y_{M}) \right]\trans = g(\vec{Y})$ can then be
used to train the GP as described in \cref{sec:gpr}, and the predictions made by the GP in
the unbounded co-domain $\tilde{\mathcal{Y}}$ can then be transformed back to the bounded
domain $\mathcal{Y}$ by applying the inverse transformation function $g\inv:
  \tilde{\mathcal{Y}} \to \mathcal{Y}$.

The training data used in this article is bounded to the domain $\mathcal{Y} =
  \mathbb{R}^{+}$~\cite{Purcell_1946}. In order to transform this data we follow the
approach taken in \cite{rickert2023high}, which extends the approach taken in
\cite{snelson2003warped}, and define a one sided piece-wise defined inverse
transformation function $g\inv$ that maps from the unbounded co-domain to the bounded
domain.

Training data between the lower bound of $y_{\mathrm{lower}} = 0$ and some lower cut-off
value $y_{\mathrm{lower, cutoff}} > y_{\mathrm{lower}}$ is transformed by means of a
simple exponential function, i.e., by $g_{\mathrm{lower}}\inv(\tilde{y}) =
  y_{\mathrm{lower}} + \exp \left( a_{\mathrm{lower}} (\tilde{y} - b_{\mathrm{lower}})
  \right)$.

Training data with $y \geq y_{\mathrm{lower, cutoff}}$ is transformed using an affine
transformation, $g_{\mathrm{linear}}\inv(\tilde{y}) = m_{\mathrm{linear}}\tilde{y} +
  b_{\mathrm{linear}}$. The steepness of the affine transformation is fixed to
$m_{\mathrm{linear}} = 1$ in accordance with \cite{rickert2023high}. It was observed that
allowing variation of $m_{\mathrm{linear}}$ only led to a reduction in the predicted
variance of the GP, and no improvement in the quality of the
prediction~\cite{rickert2023high}. This was considered as a form of overfitting. The
parameter was therefore kept constant.

The two segments are matched such that they yield a continuously differentiable function
$g\inv$. The transformation is controlled by the cut-off bound $y_{\mathrm{lower,cutoff}}$
and the position parameter of the exponential segment, $b_{\mathrm{lower}}$. The remainder
of the transformation parameters can be calculated from these control parameters. The
parameters $y_{\mathrm{lower,cutoff}}$ and $b_{\mathrm{lower}}$ are determined by means of
a hyperparameter optimization~\cite{snelson2003warped}.

\subsection{Robust design evaluation with Monte Carlo approximation}
\label{sec:mc}

Monte Carlo methods~\cite{robert1999monte,murphy2012machine} are powerful approaches that
can be used to tackle statistical problems that are difficult to solve analytically. They
generally involve a large number of random numbers from some probability distribution that
are used to perform computations. The results are aggregated and can be further used to
perform, e.g., a statistical analysis~\cite{hastie2009elements}. Monte Carlo methods can
also be applied to assess the robustness of a design for a device under given
uncertainties in the devices parameters~\cite{martin2006monte,lee2006global}.

Assume that we constructed a forward model function $f(\vec{p})$ which resembles the
device that we want to manufacture in reality. Using a Bayesian optimization (BO) method,
we have identified a model parameter $\vec{p}_{\mathrm{opt}}$ for which $f(\vec{p})$
assumes a maximum. If we were able to manufacture the device in reality, even assuming
that the model perfectly captures its characteristics and the underlying physics, we will
usually not be able to achieve the optimized model function value
$f(\vec{p}_\mathrm{opt})$. This is because typically the device parameters realized in the
manufacturing process are associated with some uncertainty and scatter around
$\vec{p}_{\mathrm{opt}}$ to a certain extent~\cite{phc_errors}.

Using a Monte Carlo approach, we can quantify the actually realized performance of the
device. For this, we assume that the parameters that are realized by the manufacturing
process scatter according to some manufacturing distribution $\mathcal{D}$ that is derived
from the uncertainties associated with the manufacturing process, and then draw a large
number of $M_{\mathrm{device}}$ random samples ${\vec{P}}_{\mathrm{device}} = \{
  {\vec{p}}_{1}, \dots, {\vec{p}}_{M_{\mathrm{device}}} \}$ from this distribution. We use
these to evaluate the devices model function, i.e., calculate ${\vec{Y}} = \{
  f({\vec{p}}_{1}), \dots, f({\vec{p}}_{M_{\mathrm{device}}})\}$. Since the samples
${\vec{P}_{\mathrm{device}}}$ are random numbers, the function values ${\vec{Y}}$ are also
random numbers that follow some statistical distribution.

As an illustrative example for such a manufacturing distribution, we choose a
$N$-dimensional multivariate normal distribution~\cite{hastie2009elements}, e.g.,
$\mathcal{D} = \mathcal{N}(\vec{p}_{\mathrm{device}}, \mat{\Sigma}_{\mathrm{manuf}})$,
with diagonal covariance matrix $\mat{\Sigma}_{\mathrm{manuf}} = \diag(\sigma_{1}^{2},
  \dots, \sigma_{N}^{2})$. In doing this, we assume that the manufacturing process is set up
to yield devices that are on average described by the parameter
$\vec{p}_{\mathrm{device}}$, but due to the assumed uncertainties $\sigma^{2}_{i}$ in the
manufacturing process the actually realized devices all scatter around the mean according
to the covariance matrix $\mat{\Sigma}_{\mathrm{manuf}}$. By assuming a diagonal
covariance matrix, we imply that the fabrication uncertainties of the individual
parameters $p_{\mathrm{device},i}$ are not correlated in any
way~\cite{hastie2009elements}.

To quantify the devices performance under manufacturing uncertainty, we can look, e.g., at
the expected value of ${\vec{Y}}$~\cite{murphy2012machine}
\begin{gather}
  \label{eq:expected_value}
  \ex{{\vec{Y}}} = \frac{1}{M_{\mathrm{device}}}
  \sum_{m=1}^{M_{\mathrm{device}}} f({\vec{p}}_{m})
\end{gather}
and its variance~\cite{murphy2012machine}
\begin{gather}
  \label{eq:variance}
  \var{{\vec{Y}}} = \frac{1}{M_{\mathrm{device}}}
  \sum_{m=1}^{M_{\mathrm{device}}} \left( f({\vec{p}}_{m}) -
  \ex{{\vec{Y}}} \right) ^{2} .
\end{gather}
For highly symmetric distributions, the expected value and variance are good measures for
the quantification of the devices performance. In some cases, however, ${\vec{Y}}$ can
show large amounts of skew. These distributions are difficult to describe using
$\ex{{\vec{Y}}}$ and $\var{{\vec{Y}}}$.
A better description can be achieved by calculating certain percentiles of the
distribution of ${\vec{Y}}$ that relate the expected performance of the device to the
expected value and variance of a well known reference distribution. Conveniently, we draw
parallels to a normal distribution $\mathcal{N}(\mu, \sigma^2)$. Here, the 50'th
percentile or median $P_{50}$ relates to the mean $\mu$ of the normal distribution and the
16'th and 84'th percentiles $P_{16}$ and $P_{84}$ to the lower standard deviation $\mu -
  \sigma$ and upper standard deviation $\mu + \sigma$, respectively.

As such, we use these percentiles to construct analogs to the lower and upper standard
deviation for the potentially skewed distribution of ${\vec{Y}}$,
\begin{align}
  \label{eq:perc_deviations}
  \sigma_{-}({\vec{Y}}) & = P_{50}({\vec{Y}}) - P_{16}({\vec{Y}}) \quad \text{and} \\
  \sigma_{+}({\vec{Y}}) & = P_{84}({\vec{Y}}) - P_{50}({\vec{Y}}) \,,
\end{align}
respectively. The median $P_{50}({\vec{Y}})$ is related to the expected value.

Expected values calculated using Monte Carlo methods get more accurate the more samples
are used. An often used estimate for the error associated with the expected value of
${\vec{Y}}$ is given as~\cite{murphy2012machine}
\begin{gather}
  \label{eq:mc_error}
  \sigma_{\mathrm{MC}} \approx \sqrt{\frac{\var{{\vec{Y}}}}{M}} \,,
\end{gather}
which tends toward zero as $M$ tends to infinity. Mathematically, \cref{eq:mc_error} is
defined only for the mean of a Monte Carlo integral. We assume that we can apply this
error measure to the percentile based approach employed by us.

Thus, using the random sample ${\vec{P}}$ to compute random forward model function values
${\vec{Y}}$, we can give an estimate for the realizable performance of the device as
\begin{gather}
  \label{eq:expected_performance}
  \overline{f} =
  \left( P_{50}({\vec{Y}}) \pm \sigma_{\mathrm{MC}}
  \right)_{\sigma_{-}({\vec{Y}})}^{\sigma_{+}({\vec{Y}})} \,.
\end{gather}

This procedure can be incorporated into an (expensive) optimziation procedure if the mean
of the manufacturing distribution, $\vec{p}_{\mathrm{device}}$, is iteratively chosen by
an optimization method, e.g., by the BO method described earlier. The expected improvement
criterion would then select points in the parameter space where the median
$P_{50}({\vec{Y}})$ would be optimized.

\section{Robust design optimization procedure}
\label{sec:procedure}

While being conceptionally relatively simple, robust optimization of the model function
$f(\vec{p})$ directly with the approach outlined in \cref{sec:mc} is still difficult from
a computational resources standpoint, since it involves evaluating the expensive model
function a large number of times. Here, we are dealing with a few issues that compound
themselves into a very challenging task.

On one hand the size of optimization domain has to be considered. Without any prior
information for the model parameters $\vec{p}$ it is difficult to motivate a particular
choice of limited parameter range. Instead, it is desirable to investigate large intervals
in each of the problems parameter dimensions. Performing this in an unrobust fashion is
already a challenging task~\cite{Weise2009}.

On the other hand, in order to obtain a robust estimate for a given parameter the model
function $f(\vec{p})$ has to be evaluated for many points in the vicinity of $\vec{p}$, as
detailed in \cref{sec:mc}. This increases the computational cost for each considered point
in the parameter space by a factor of $M_{\mathrm{device}}$.

The procedure presented in the following aims to resolve these issues. It is centered
around cheap-to-evaluate Gaussian process surrogate models. This helps to efficiently
manage the computational burden imposed by a robust design optimization. The predictions
of this GP surrogate are used during the application of an iterative Monte Carlo sampling
technique detailed in \cref{alg:mc_int}. In it, we go through a series of five steps in
two successive passes and at the end obtain a parameter candidate that is predicted to be
robust against some assumed manufacturing uncertainties.
\begin{algorithm*}[h]
  \caption{An iterative Monte Carlo sampling procedure to determine a robust device
    performance estimate, by obtaining predictions from a trained GP surrogate at
    positions drawn from some manufacturing distribution $\mathcal{D}$.}
  \label{alg:mc_int}
  \begin{algorithmic}
    \Function{RobustEstimate}{Predicted GP mean $\overline{y}(\vec{p})$
      (\cref{eq:gp_pred_mean}), predicted GP variance $\sigma^{2}(\vec{p})$
      (\cref{eq:gp_pred_var}), manufacturing distribution $\mathcal{D}$}

    \State $\sigma_{\mathrm{rel,MC}} = \infty$ \Comment
    \parbox[t]{.57\linewidth}{Initialize the relative MC error.}

    \State $N_{\mathrm{tot}} = 0$ and $N_{\mathrm{min}}$ = \num{50000} \Comment
    \parbox[t]{.57\linewidth}{Initialize the number of samples drawn.}

    \State ${\vec{Y}}_{\mathrm{tot}} = \emptyset$ and ${\vec{S}}_{\mathrm{tot}} =
      \emptyset$ \Comment \parbox[t]{.57\linewidth}{Initialize the sets of GP mean and
      variance observations.}

    \While{$( \sigma_{\mathrm{rel,MC}} \geq \num{e-3}$ \textbf{and} $N_{\mathrm{tot}} <
        N_{\mathrm{min}} )$}

    \State ${\vec{P}} \sim \mathcal{D}$ and  $N_{\mathrm{tot}} = N_{\mathrm{tot}} +
      \num{1000}$ \Comment \parbox[t]{.57\linewidth}{Draw \num{1000} random samples from
      manufacturing distribution.}

    \State ${\vec{Y}} = \overline{y}({\vec{P}})$ and ${\vec{S}} = \sigma^{2}({\vec{P}})$
    \Comment \parbox[t]{.57\linewidth}{Use samples to calculate predicted GP mean and
      variance values (\cref{eq:gp_pred_mean,eq:gp_pred_var}).}

    \State ${\vec{Y}}_{\mathrm{tot}} = {\vec{Y}}_{\mathrm{tot}} \cup {\vec{Y}}$ and
    ${\vec{S}}_{\mathrm{tot}} = {\vec{S}}_{\mathrm{tot}} \cup {\vec{S}}$ \Comment
    \parbox[t]{.57\linewidth}{Collect mean and variance observations.}

    \State $\sigma_{\mathrm{rel,MC}} = \sigma_{\mathrm{MC}}({\vec{Y}}_{\mathrm{tot}}) /
      P_{50}({\vec{Y}}_{\mathrm{tot}})$ \Comment \parbox[t]{.57\linewidth}{Check
      convergence of MC sampling (using \cref{eq:mc_error}).}

    \EndWhile

    \State $\sigma_{\mathrm{GP}}^{2} = P_{50}({\vec{S}}_{\mathrm{tot}})$ \Comment
    \parbox[t]{.57\linewidth}{Use variance observations to estimate error introduced by
      GPs.}

    \State $\sigma_{\mathrm{Median}} = \sqrt{\sigma_{\mathrm{GP}}^{2} +
        \sigma_{\mathrm{MC}}^{2}}$ \Comment \parbox[t]{.57\linewidth}{Calculate
      uncertainty for the predicted device performance (using \cref{eq:mc_error}.)}

    \State $\sigma_{-}({\vec{Y}}_{\mathrm{tot}}) = P_{50}({\vec{Y}}_{\mathrm{tot}}) -
      P_{16}({\vec{Y}}_{\mathrm{tot}})$ \Comment \parbox[t]{.57\linewidth}{Use mean
      observations to calculate lower \dots} \State $\sigma_{+}({\vec{Y}}_{\mathrm{tot}})
      = P_{84}({\vec{Y}}_{\mathrm{tot}}) - P_{50}({\vec{Y}}_{\mathrm{tot}})$ \Comment
    \parbox[t]{.57\linewidth}{\dots and upper standard deviation for the predicted
      device performance.}

    \State \textbf{return} $\left( P_{50}({\vec{Y}}_{\mathrm{tot}}) \pm
      \sigma_{\mathrm{Median}}
      \right)^{\sigma_{+}({\vec{Y}}_{\mathrm{tot}})}_{\sigma_{-}({\vec{Y}}_{\mathrm{tot}})}$
    \Comment \parbox[t]{.57\linewidth}{Return robust device performance estimate.}

    \EndFunction
  \end{algorithmic}
\end{algorithm*}

In the first step, training data is generated by evaluating the considered model function
$f(\vec{p})$ at many different positions across the parameter space. In step two, a GP
surrogate model for bounded data is fit to the training data. In step three, the trained
GP surrogate is employed to generate robust estimates of the model function output at many
positions in the computational domain. The quality of these estimates scales inversely
with the dimensionality of the parameter space and the size of the investigated domain. In
step four, the best robust estimate(s) found in step three is (are) converged using a
Bayesian optimizer. Finally, in step five, the converged candidates are inspected and
their robustness is verified using the expensive forward model function $f(\vec{p})$.

Using two distinct passes helps dealing with the size of the optimization domain. In the
first pass, a very large domain is considered in a rather coarse fashion. The parameter
candidate generated here only serves to define a much smaller domain to be investigated in
the second pass. The parameter candidate generated by the second pass finally is predicted
to be robust against the assumed manufacturing uncertainties.

There is, however, no guarantee that we might overlook a very good candidate, since the
parameter space in the first pass can be very large when compared to the length scales on
which the forward model function $f(\vec{p})$ varies. This is a computational cost
problem, that can be resolved, e.g., by partitioning the computational domain and then
investigating each of these partitions in a more refined way. This approach does, however,
not scale that well for high dimensional problems, which is a direct manifestation of the
curse of dimensionality~\cite{kuo2005lifting}.

In addition to the optimized forward model function $f(\vec{p})$, the procedure also
requires an optimization domain and a set of assumed manufacturing uncertainties. The
uncertainties are used to construct manufacturing distributions employed during the robust
optimization. Here, we limit our considerations to simple multivariate normal
distributions~\cite{hastie2009elements} with diagonal covariance matrix
$\mat{\Sigma}_{\mathrm{manuf}}$. The approach is easily extended to more complex
distributions.

The individual steps are discussed in more detail in the following.

\subsection{Step 1: Generating training data}
\label{sec:procedure_training_data}

Given a domain $\mathcal{Q}_{\mathrm{train}} \subset \mathcal{X}$, a forward model
function $f(\vec{p})$ that is to be optimized in a robust fashion, and the assumed
manufacturing uncertainties $\sigma_{d_{i}}$ for each parameter, we generate a large
number of training data points in $\mathcal{Q}_{\mathrm{train}}$. The manufacturing
uncertainty imposes minimum-size-requirement on the domain $\mathcal{Q}_{\mathrm{train}}$.
It must be possible to position a constructed manufacturing normal distribution
$\mathcal{D}$ in a way that, when drawing many random samples from it, a very large
portion of these samples are in $\mathcal{Q}_{\mathrm{train}}$. We will elaborate on this
in \cref{sec:procedure_batch_robust}.

Samples $\vec{P}_{\mathrm{train}}$ are drawn from a Sobol
sequence~\cite{sobol1967distribution,joe2008constructing} scaled to
$\mathcal{Q}_{\mathrm{train}}$ and used to evaluate $f(\vec{p})$, thereby generating the
training data $\vec{Y}_{\mathrm{train}} = f(\vec{P}_{\mathrm{train}})$. Sobol sequences
are low-discrepancy sequences of points in a $N$-dimensional space that cover the space
somewhat evenly, provided that the number of used samples equals to a natural power of
two~\cite{owen2020dropping}. Because of the costs associated with the evaluation of the
forward model function and because the performance of GPR decreases drastically if a very
large number of samples are used to train it~\cite{rasmussen2003gaussian}, we draw $2^{12}
  = \num{4096}$ samples from the Sobol sequence. Depending on the dimensionality and size of
the parameter space of the model, this choice may result in a sparse sampling of the
space. For the dimensionalities considered by us, also, e.g., in~\cite{rickert2023high},
this number was demonstrated to be sufficient. The upper limit for this number is
constrained computationally, i.e., a surrogate with $\num{8192}$ samples may not be
tractable anymore.

Some filtering of $\vec{Y}_{\mathrm{train}}$ for the next step can be beneficial, since a
GP generally assumes that the data used to train it could have been drawn from \emph{some}
GP, i.e., it is normally distributed~\cite{garnett_bayesoptbook_2023}. As such, extreme
outliers in $\vec{Y}_{\mathrm{train}}$ should be removed to assure that the predictions of
the trained GP are a good approximation of the function $f(\vec{p})$.

\subsection{Step 2: Training the GP surrogate}
\label{sec:procedure_training_gp}

The (potentially filtered) bounded training data $\vec{Y}_{\mathrm{train}}$ and associated
samples $\vec{P}_{\mathrm{train}}$ are used to train the GP surrogate model for bounded
data described in \cref{sec:bounded_gpr}. This involves finding all the hyperparameters
for the transformation function $g$ that is used to transform the bounded training data to
an unbounded co-domain for training the GP, and transforming the predictions of the GP
back to the bounded domain by applying $g\inv$. Depending on the number of training data
points and on the size of the domain $\mathcal{Q}_{\mathrm{train}}$, this surrogate can be
a coarse or fine approximation of the forward model function $f(\vec{p})$ on the domain
$\mathcal{Q}_{\mathrm{train}}$.

\subsection{Step 3: Batch calculation of robust estimates}
\label{sec:procedure_batch_robust}

A large number of $M_{\mathrm{eval}}$ samples $\vec{P}_{\mathrm{eval}} = \left[
  \vec{p}_{1}, \dots, \vec{p}_{M_{\mathrm{eval}}} \right]$ are drawn from a Sobol sequence
that is scaled to the domain $\mathcal{Q}_{\mathrm{eval}} \subset
  \mathcal{Q}_{\mathrm{train}}$. These are used to construct various manufacturing normal
distributions $\mathcal{D}_{i} = \mathcal{N}(\vec{p}_{i},
  \mat{\Sigma}_{\mathrm{manuf}})$. In conjunction with the trained GP surrogate on
$\mathcal{Q}_{\mathrm{train}}$, they are used to generate robust estimates of the
forward model function $f(\vec{p})$, by applying \cref{alg:mc_int}.

The domain $\mathcal{Q}_{\mathrm{eval}}$ is reduced in size by three standard deviations
in each direction and each parameter when compared to $\mathcal{Q}_{\mathrm{train}}$. This
is done to assure that the samples drawn from the manufacturing distribution
$\mathcal{D}_{i}$ during the application of \cref{alg:mc_int} are located in regions where
the predictions of the GP surrogate are informed by training data. Consider a parameter
$\vec{p}_{\mathrm{corner}} \in \mathcal{Q}_{\mathrm{eval}}$ which sits in a corner of the
domain $\mathcal{Q}_{\mathrm{eval}}$. Determining a robust estimate for this parameter
means drawing a large number of samples from the normal distribution
$\mathcal{N}(\vec{p}_{\mathrm{corner}}, \mat{\Sigma}_{\mathrm{manuf}})$. By the properties
of normal distributions, approximately \SI{99.7}{\percent} of all drawn samples will be
contained in a hypersphere with a radius of three standard deviations defined by the
covariance matrix $\mat{\Sigma}_{\mathrm{manuf}}$~\cite{schervish2014probability}. Since
the training domain $\mathcal{Q}_{\mathrm{train}}$ of the GP surrogate is three standard
deviations larger, this means that more than \SI{99.7}{\percent} of all predictions will
be informed by training data. Moving the center of the normal distribution outside of
$\mathcal{Q}_{\mathrm{eval}}$ reduces this fraction. It is in principle possible to use
any manufacturing distribution that can be reasonably motivated, e.g. also a bounded
distribution if one of the parameters is e.g. a radius that can not be negative. This
includes the possibility to encode correlations between model parameters by, e.g., using a
covariance matrix $\mat{\Sigma}_{\mathrm{manuf}}$ with off-diagonal elements. The
performance of the scheme is unlikely to suffer from this, since it is typically possible
to generate a relevant number of samples from various types of distributions.

\cref{alg:mc_int} extends the sampling procedure described in \cref{sec:mc} by one aspect.
Instead of sampling the forward model function $f(\vec{p})$ directly, a GP surrogate model
of the function is sampled. GP surrogates are always associated with some uncertainty in
the prediction, quanfitied by the predicted variance, \cref{eq:gp_pred_var}. This variance
prediction can be utilized to extend \cref{eq:expected_performance} and to give an
estimate for the uncertainty of the predicted robust estimate, too. For this, the random
samples ${\vec{P}}$ drawn during application of \cref{alg:mc_int} are used to generate
predicted variances ${\vec{S}} = \sigma^{2}({\vec{P}})$. The median of these variance
samples $\sigma_{\mathrm{GP}}^{2} = P_{50}({\vec{S}})$ is used together with the Monte
Carlo error, \cref{eq:mc_error}, to give a combined predicted uncertainty as
\begin{gather}
  \label{eq:combined_uncertainty}
  \sigma_{\mathrm{Median}} = \sqrt{\sigma_{\mathrm{GP}}^{2} +
    \sigma_{\mathrm{MC}}^{2}} \,.
\end{gather}
The magnitude of the median of the variance samples will dominate the Monte Carlo error.
We believe it to be a reasonable assumption that this will also be the case for the
percentile based approach taken by us. Using this we can give the estimate of the
realizable device performance as
\begin{gather}
  \label{eq:gp_expected_performance}
  \overline{f} =
  \left( P_{50}({\vec{Y}}) \pm \sigma_{\mathrm{Median}}
  \right)_{\sigma_{-}({\vec{Y}})}^{\sigma_{+}({\vec{Y}})} \,.
\end{gather}

Conventionally, one would not use a batch evaluation of the robust estimate, but rather
employ some form of optimization method to scan the parameter space. We have, however,
observed that even optimization methods that are designed for global optimization (we used
BO) had issues coping with the size of the investigated domain in the first pass.

This is associated with two issues in particular. First, the BO method tends to exploit
promising values rather than exploring the parameter space. While this approach may work
in the second pass through the pipeline, in the first pass, it spends a large number of
evaluations that could much better be spent exploring the parameter space. And second,
this exploitation of promising values was tied to collecting many training data points. At
each BO iteration a matrix has to be inverted that scales with the number of training data
points~\cite{rasmussen2003gaussian,garnett_bayesoptbook_2023}. As such, the BO method
becomes very slow after many samples have been investigated.

These issues could in principle be resolved by using a conventional optimization method,
such as L-BFGS-B~\cite{byrd1995limited}, albeit at the expense of the global optimization
property of the method. To therefore reliably cover a wide area in the parameter space one
could employ a random multi start approach~\cite{dick2014many}. However, we deemed it more
efficient to sample well known locations in the parameter space directly.

The results obtained using the batch approach are sorted by their predicted performance.
Good results that are in very close proximity to other good results are filtered, such
that only the best results from a cluster is kept. A limited number of best filtered
results are then investigated in the next step.

\subsection{Step 4: Converging robust design candidates}
\label{sec:procedure_converging_candidates}

The filtered results were each used as starting point for a Bayesian optimizer, by
pre-training the optimization method on them. This pre-training has nudged the optimizer
into exploiting the area surrounding the filtered result. The BO method iteratively chose
the mean $\vec{p}$ of the manufacturing normal distribution $\mathcal{D} =
  \mathcal{N}(\vec{p}, \mat{\Sigma}_{\mathrm{manuf}})$ used during the application of
\cref{alg:mc_int} on the trained GP surrogate. As in the previous step, the domain
$\mathcal{Q}_{\mathrm{eval}}$ is used as optimization domain.

\subsection{Step 5: Inspection and verification of converged candidates}
\label{sec:procedure_inspection}

This step is primarily important for the first pass, as it is used to determine ranges in
the parameter space to investigate more closely. It is not strictly necessary for the
second pass, but can still be interesting to verify the predictions of the GP surrogates
and check that the results are sensible.

The predicted results for the converged parameters found in the previous step are verified
using the expensive forward model function $f(\vec{p})$. A limited number of random
samples ${\vec{P}}_{\mathrm{test}}$ are drawn from the manufacturing distribution centered
on the converged robust design candidates and used to evaluate $f(\vec{p})$. The median of
the results is calculated. In order to select a parameter region for the second pass, the
best median forward model result is selected.

Note that the number of samples investigated in the second pass should be large enough
such that the Monte Carlo error associated with the final converged results does not
dominate these~\cite{robert1999monte,murphy2012machine}.

\subsection*{A remark}
\label{sec:closing_remark}

For the sake of clarity, we have included only the basics of the five steps in this
Section. Details of the complete two pass approach will be explained alongside an
application example in
\cref{sec:results}

\section{Application}
\subsection{Model of a photonic nanobeam cavity}
\label{sec:example}

\begin{figure*}[h]
  \includegraphics{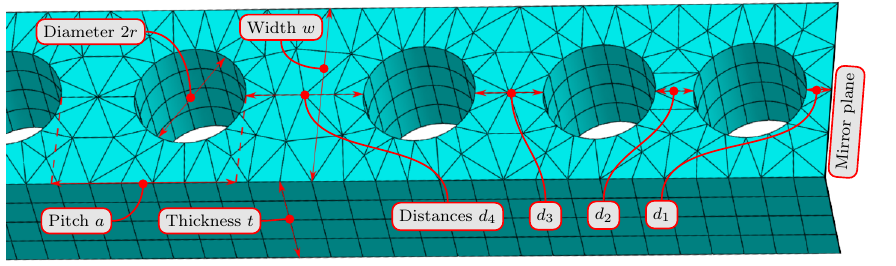}
  \caption{\label{fig:full_inner_cavity} An exemplary parameterization of the cavity. All
    cavity parameters are illustrated by means of red arrows. The hole diameters $2r$, as
    well as cavity width $w$, pitch $a$, and thickness $t$ are determined by optimizing
    the crystal unit cell. The distances $d_{1}$ to $d_{4}$ are chosen to yield a cavity
    whose Purcell factor $F_{\mathrm{P}}$ is robust under a given fabrication uncertainty
    for the distances.}
\end{figure*}
To demonstrate the method, we optimize a hole based photonic crystal nanobeam
cavity~\cite{joannopoulos1997photonic,Deotare2012} in a robust way. The cavity is an open
system, i.e., it is described by
resonances~\cite{Lalanne_QNMReview_2018,Kristensen_QNM_2020}. The maximized target
quantity is the Purcell enhancement $F_{\mathrm{P}}$ of the cavity, which is given by the
quotient of the total emission of a dipole embedded in the cavity and the dipole emission
in homogeneous bulk material~\cite{Purcell_1946,Sauvan_QNMexpansionPurcell_2013}. The
dipole emits at a wavelength of \SI{1550}{\nano\meter}, i.e., in the telecom C-band
wavelength~\cite{muller2018quantum}. The model is set up and solved using the Maxwell
solver JCMsuite which is based on the finite-element
method~\cite{Pomplum_NanoopticFEM_2007}. In principle however any method that can be used
to calculate rigorous solutions to Maxwells equations can be used instead.

The photonic crystal nanobeam cavity is constructed by repeating a finite number of unit
cells periodically along a single dimension. The unit cells are modeled as a rectangular
block of silicon with thickness $t$, width $w$, and length or pitch $a$. A hole through
the block with diameter $2r$ is positioned in the center of the $a$-$w$
plane~\cite{joannopoulos1997photonic}. The band structure of the photonic crystal contains
some fundamental dielectric band $\lambda_{\mathrm{diel}}$ and a next higher air band
$\lambda_{\mathrm{air}}$, which can form a band gap~\cite{joannopoulos1997photonic}. In
order to create a cavity mode within this crystal, the location
$\overline{\lambda}_{\mathrm{gap}}$ and size $\Delta \lambda_{\mathrm{gap}}$ of this band
gap within the band structure must be controlled~\cite{joannopoulos1997photonic}.

We can then create a resonant cavity by introducing a defect at some position of the
crystal, which breaks up the periodicity of the crystal. A very simple way of doing this
is by, e.g., varying the distance between two of the holes. This variation modifies the
amount of material in the crystal at that position and therefore also the band structure,
which can allow for the emergence of a resonance mode in the region of the defect.
Ideally, due to the different band structure, this resonance mode can not propagate inside
the periodic region. The periodic region acts thus as a mirror for the resonance mode. A
more thorough treatment of the topic can be found in the
literature~\cite{joannopoulos1997photonic}.

The task at hand is two fold. First, the parameters of the periodic unit cell, i.e., the
pitch $a$, the width $w$, the thickness $t$, and the diameter $2r$, must be chosen such
that the periodic region serves as an effective mirror for a desired resonance mode, by
disallowing its propagation entirely. And, second, degrees of freedom have to be
introduced to parameterize the crystal defect, and these have to be chosen such that the
desired resonance mode emerges in the defect region.

The parameters of the unit cell are chosen to create a photonic crystal with a very large
band gap centered around (or close to) the desired wavelength of \SI{1550}{\nano\meter}.
For this, a Bayesian optimizer is used to minimize the objective function
\begin{gather}
  \label{eq:bs_target_function}
  b(\vec{q}) = -1 \cdot \frac{\Delta
  \lambda_{\mathrm{gap}}(\vec{q})}{| \overline{\lambda}_{\mathrm{gap}}(\vec{q})
  - \SI{1550}{\nano\meter} |} \,, \\
  \text{with} \quad \Delta \lambda_{\mathrm{gap}}(\vec{q}) =
  \lambda_{\mathrm{diel}} - \lambda_{\mathrm{air}}  \nonumber \\
  \text{and} \quad \overline{\lambda}_{\mathrm{gap}}(\vec{q}) = \left(
  \lambda_{\mathrm{diel}} + \lambda_{\mathrm{air}} \right) / 2 \,, \nonumber \\
  \text{where} \quad \vec{q} = \left[ a, w, t, 2r \right]\trans \,. \nonumber
\end{gather}
The fundamental dielectric and next higher air mode, $\lambda_{\mathrm{diel}}$ and
$\lambda_{\mathrm{air}}$, respectively, are calculated by solving resonance problems. The
objective functions $b(\vec{q})$ output becomes very small if the gap in the band
structure $\Delta \lambda_{\mathrm{gap}}(\vec{q})$ becomes very large and the center of
the gap in the band structure $\overline{\lambda}_{\mathrm{gap}}(\vec{q})$ is very close
to the target wavelength of \SI{1550}{\nano\meter}. We find that a unit cell with pitch $a
  = \SI{560}{\nano\meter}$, width $w = \SI{620}{\nano\meter}$, thickness $t =
  \SI{395}{\nano\meter}$, and a hole diameter of $2r = \SI{336}{\nano\meter}$ has a band gap
of $\Delta\lambda_{\mathrm{gap}} \approx \SI{214}{\nano\meter}$, centered around a
wavelength of $\overline{\lambda}_{\mathrm{gap}} \approx \SI{1554}{\nano\meter}$. The
fundamental dielectric mode has a wavelength of $\lambda_{\mathrm{diel}} \approx
  \SI{1661}{\nano\meter}$ and the next higher air mode a wavelength of
$\lambda_{\mathrm{air}} \approx \SI{1447}{\nano\meter}$.

The photonic crystal nanobeam cavity is created by periodically repeating 13 unit cells in
the computational domain. We further consider a mirror symmetric setup, i.e., we treat one
side of the computational domain as a mirror plane. This effectively doubles the number of
unit cells in the device. For the introduction and variation of the defect in the periodic
structure, we consider the mirror plane, i.e., the center of the photonic crystal. Here,
we allow the distances between the four innermost holes $d_{4}$, $d_{3}$, and $d_{2}$, as
well as the distance from the first hole to the mirror plane, $d_{1}$, to vary. As such,
we consider the four distance parameters $\vec{p} = \left[ d_{1}, d_{2}, d_{3}, d_{4}
    \right]$ for parameterization of the defect. The cavity consists of 10 mirror cells and 3
cavity cells on each side of the mirror plane. An excerpt of the geometrical layout of the
cavity is depicted in \cref{fig:full_inner_cavity}.

The Purcell enhancement $F_{\mathrm{P}}$ of this structure is calculated by means of
scattering simulations~\cite{Pomplum_NanoopticFEM_2007}, in which a dipole with an
emission wavelength of $\lambda_{\mathrm{dipole}} = \SI{1550}{\nano\meter}$ is placed in
the mirror plane, in the center of the waveguide material.

\subsection{Robust design optimization of the nanobeam cavity}
\label{sec:results}

We have performed a robust optimization of the photonic crystal nanobeam cavity described
in \cref{sec:example}. During the optimization, we considered the cavity parameters
$d_{1}$ to $d_{4}$ on the domain given in \cref{tab:full_space}.
\begin{table}[h]
  \centering
  \begin{tabular}{cc}
    Design parameter & Parameter range                            \\
    \toprule
    $d_{1}$          & $\left[ 56, 616 \right]\,\si{\nano\meter}$ \\
    $d_{2}$          & $\left[ 56, 616 \right]\,\si{\nano\meter}$ \\
    $d_{3}$          & $\left[ 56, 616 \right]\,\si{\nano\meter}$ \\
    $d_{4}$          & $\left[ 56, 616 \right]\,\si{\nano\meter}$ \\
    \bottomrule
  \end{tabular}
  \caption{\label{tab:full_space} The complete investigated design space. This domain
  $\mathcal{Q}_{\mathrm{train,first}}$ is used as a basis for the first pass of the
  optimization process described in \cref{fig:schematic}.}
\end{table}
For each of the $d_{i}$, we assumed a manufacturing uncertainty of $\sigma_{d_{i}} =
  \SI{16.8}{\nano\meter}$---note that this equals \SI{3}{\percent} of the pitch $a$ of a
unit cell---and no correlation between the $d_{i}$. This means that the combined
manufacturing uncertainty can be modeled by a diagonal covariance matrix
$\mat{\Sigma}_{\mathrm{manuf}} = \diag(\sigma_{d}^2, \sigma_{d}^2, \sigma_{d}^2,
  \sigma_{d}^2)$. The unit cell parameters $a$, $w$, $t$, and $2r$ were kept constant and,
for the sake of this demonstration, assumed not to be associated with any manufacturing
uncertainty.

The following analysis closely follows the description given in \cref{sec:procedure} and
the schematic in \cref{fig:schematic}. First, in a coarse pass, a candidate region was
identified which shows a promising range of Purcell enhancements that allow for a robust
solution under the assumed manufacturing uncertainties. This candidate region was then
investigated more closely in a narrow and more refined second pass. Here, a parameter set
was determined which maximizes the Purcell enhancement in a robust fashion such that
moderate variations of the parameters yield similar Purcell enhancements. Such a result is
much easier to realize in a manufacturing environment.

We used transformation based GPs since we expected that a very large number of training
data values were very close to \num{0}, because the Purcell enhancement can be assumed to
be very small for large portions of the parameter space.

\subsubsection{First pass: candidate selection}
\label{sec:results_first_pass}

A set of \num{4096} parameters $\vec{P}_{\mathrm{train,first}}$ was drawn from a Sobol
sequence for the domain $\mathcal{Q}_{\mathrm{train,first}}$ given in
\cref{tab:full_space}. These were used to generate training data
$\vec{Y}_{\mathrm{train,first}} = f(\vec{P}_{\mathrm{train,first}})$ from the FEM model
for the photonic crystal nanobeam cavity. The training data was inspected and data points
with large values with $F_{\mathrm{P}} > 50$ were removed. This applied to only two data
points. The reason for this filtering is that a GP assumes training data that could in
principle have been drawn from \emph{some} GP, i.e., is normally distributed. Assuming
that $\vec{Y}_{\mathrm{train,first}}$ roughly follows a normal distribution, values larger
than $F_{\mathrm{P}} = 50$ appeared uncharacteristically large. We will return to the
largest value just removed from the data set in \cref{sec:comparing_to_naive}.

The remaining \num{4094} data points were used to train a transformation based GP
surrogate of the forward model function on the domain
$\mathcal{Q}_{\mathrm{train,first}}$. Note that, due to the number of training points and
the volume of the domain, this GP is a relatively coarse approximation of $f(\vec{p})$.

The trained GP was used to generate a large set of robust estimates of the Purcell
enhancement. These robust estimates were calculated by applying \cref{alg:mc_int} for
different manufacturing distributions $\mathcal{D}$. In accordance with the assumed
manufacturing uncertainties $\sigma_{d_{i}}$, these were all normal distributions, i.e.,
$\mathcal{D} = \mathcal{N}(\vec{p}, \mat{\Sigma}_{\mathrm{manuf}})$, which differed only
in the mean $\vec{p}$ of the distributions. To quickly and efficiently cover the parameter
space, we drew \num{4096} samples to be used as mean values $\vec{p}$ from a Sobol
sequence in the domain $\mathcal{Q}_{\mathrm{eval,first}} \subset
  \mathcal{Q}_{\mathrm{train,first}}$. The robust estimates were sorted by magnitude. Good
values in close proximity (we used $d_{ij} = \lVert \vec{p}_{i} - \vec{p}_{j} \rVert <
  0.25$ as criterion) were filtered such that only the largest robust estimate was kept.

The six largest filtered results were each converged with a Bayesian optimizer. The
objective function optimized was \cref{alg:mc_int}, where the mean $\vec{p}$ of the
manufacturing distribution was iteratively chosen to maximize the robust Purcell
enhancement. Each of the six filtered results from the previous step were individually
used as starting points for the optimization. It was found that two of the six candidates
consistently converged into the same optimum, leaving five distinct candidates.

To identify the most promising parameter to inspect in the second pass, \num{64} samples
were drawn from appropriate manufacturing distributions $\mathcal{D}$ centered around each
of the five candidates. These were used to evaluate the FEM model. We calculated the
median of the calculated Purcell enhancements and selected the candidate with the highest
value. Here, the parameter $\vec{p}_{\mathrm{opt,first}} = \left[ 428.6, 282.5, 369.5, 253
    \right]\,\si{\nano\meter}$ was selected for a more thorough inspection in the second pass.
Taking the very coarse nature of the surrogate into account, we have seen a reasonable
agreement between the results obtained using the predictions made by the GP surrogate
(median of \num{9.4}) and the FEM model (median of \num{2.3}).

\subsubsection{Second pass: robust optimization of candidates}
\label{sec:results_second_pass}

The second, more refined pass was centered on the parameter $\vec{p}_{\mathrm{opt,first}}
  = \left[ 428.6, 282.5, 369.5, 253 \right]\,\si{\nano\meter}$ which has been identified
in the first pass. A set of \num{4096} training data points were generated on a domain
spanning a total of ten standard deviations in each dimension, i.e., the domain
$\mathcal{Q}_{\mathrm{train,second}}$ listed in \cref{tab:narrow_space}.
\begin{table}[h]
  \centering
  \begin{tabular}{cc}
    Design parameter & Parameter range                                 \\
    \toprule
    $d_{1}$          & $\left[ 344.6, 512.6 \right]\,\si{\nano\meter}$ \\
    $d_{2}$          & $\left[ 198.5, 366.5 \right]\,\si{\nano\meter}$ \\
    $d_{3}$          & $\left[ 285.5, 453.5 \right]\,\si{\nano\meter}$ \\
    $d_{4}$          & $\left[ 169, 337 \right]\,\si{\nano\meter}$     \\
    \bottomrule
  \end{tabular}
  \caption{\label{tab:narrow_space} After a promising design candidate was found at
  $\vec{p}_{\mathrm{opt,first}} = \left[ 428.6, 282.5, 369.5, 253
      \right]\,\si{\nano\meter}$, the model was investigated more closely in this domain
  $\mathcal{Q}_{\mathrm{train,second}}$ centered around the value. This domain was used as
  a basis for the second pass of the optimization process described in
  \cref{fig:schematic}. This is a ten standard deviation hypercube centered around
  $\vec{p}_{\mathrm{opt,first}}$.}
\end{table}
\begin{figure*}[h]
  \includegraphics{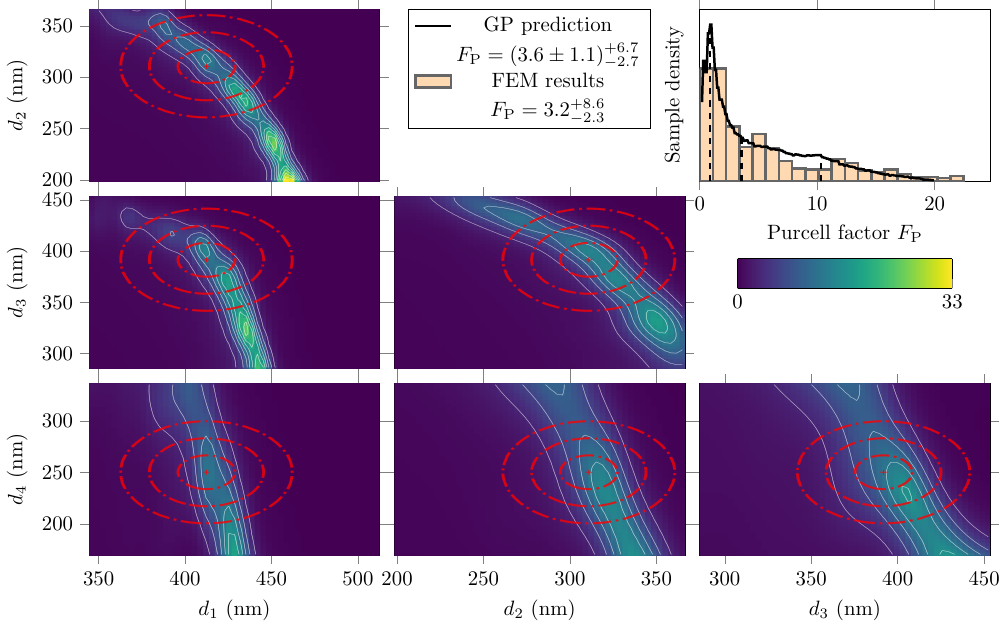}
  \caption{\label{fig:three_percent} The function value landscape around the robust design
  parameter $\vec{p}_{\mathrm{robust}} = \left[ 412, 311.5, 392, 250.3
      \right]\,\si{\nano\meter}$ (marked with a red cross). The red ellipses around the red
  cross indicate (from smallest to largest) the $1\sigma$, $2\sigma$, and $3\sigma$
  percentiles of the manufacturing distribution, which was assumed to be a multivariate
  normal distribution. The covariance matrix of this multivariate normal distribution was
  chosen to be $\mat{\Sigma}_{\mathrm{manuf}} = \diag(\sigma_{d}^{2}, \sigma_{d}^{2},
    \sigma_{d}^{2}, \sigma_{d}^{2})$ with $\sigma_{d} = \SI{16.8}{\nano\meter}$. Applying
  \cref{alg:mc_int} in conjunction with a trained surrogate model allowed us to
  efficiently determine that a manufacturing process that produces devices according to a
  normal distribution with $\mathcal{N}(\vec{p}_{\mathrm{robust}},
    \mat{\Sigma}_{\mathrm{manuf}})$ yields devices with an expected Purcell enhancement
  $\overline{F}_{\mathrm{{P}}} = (3.6 \pm 1.1)_{-2.7}^{+6.6}$. Verification using the FEM
  model showed that $\overline{F}_{\mathrm{P}}= 3.2_{-2.3}^{+8.6}$. Both results are shown
  in the top right sub figure.}
\end{figure*}
These $\vec{P}_{\mathrm{train,second}}$ were again drawn from a Sobol sequence and used to
evaluate the FEM model with, thereby generating $\vec{Y}_{\mathrm{train,second}} =
  f(\vec{P}_{\mathrm{train,second}})$. These results were inspected to detect extreme
outliers similar to the first pass.

Because none were found, all \num{4096} data points were used for training the
transformation based GP. The training data spans a much smaller domain, as such the
resulting surrogate can be considered a much more accurate approximation of the FEM model
$f(\vec{p})$.

A large set of robust estimates of the Purcell enhancements were generated by using
\cref{alg:mc_int} in conjunction with the finer surrogate trained. Similar to the first
pass, \num{4096} different normal distributions centered on points $\vec{p} \in
  \mathcal{Q}_{\mathrm{eval,second}} \subset \mathcal{Q}_{\mathrm{train,second}}$ were used
to generate \num{4096} robust estimates. The best result was finally converged using BO.

We have determined the parameter $\vec{p}_{\mathrm{robust}} = \left[ 412, 311.5, 392,
    250.3 \right]\,\si{\nano\meter}$ as robust against uncertainty in the manufacturing
process, assuming a manufacturing uncertainty of $\sigma_{d_{i}} =
  \SI{16.8}{\nano\meter}$ for each parameter $d_{i}$. The function value landscape
around $\vec{p}_{\mathrm{robust}}$, as determined from the GP surrogate, is shown in
\cref{fig:three_percent}. The small red cross shows $\vec{p}_{\mathrm{robust}}$, and
the red ellipses around it denote the $1\sigma$, $2\sigma$, and $3\sigma$ percentiles
of a normal distribution. Applying \cref{alg:mc_int} with $\mathcal{D} =
  \mathcal{N}(\vec{p}_{\mathrm{robust}}, \mat{\Sigma}_{\mathrm{manuf}})$ together with
the fine surrogate yields a robust estimate for the Purcell enhancement of
$\overline{F}_{\mathrm{P}} = (3.6 \pm 1.1)_{-2.7}^{+6.6}$. We verified this result by
evaluating the FEM model with \num{512} random samples drawn from $\mathcal{D}$,
resulting in a slightly smaller value of $\overline{F}_{\mathrm{P}}=
  3.2_{-2.3}^{+8.6}$. The distributions of the GP predictions and the FEM results are
found in the top right subplot in \cref{fig:three_percent}.

In \cref{secapp:variation} we have performed this second pass with an assumed
manufacturing uncertainty of $\sigma_{d_{i}} = \SI{11.2}{\nano\meter}$.

\begin{figure*}[h]
  \includegraphics{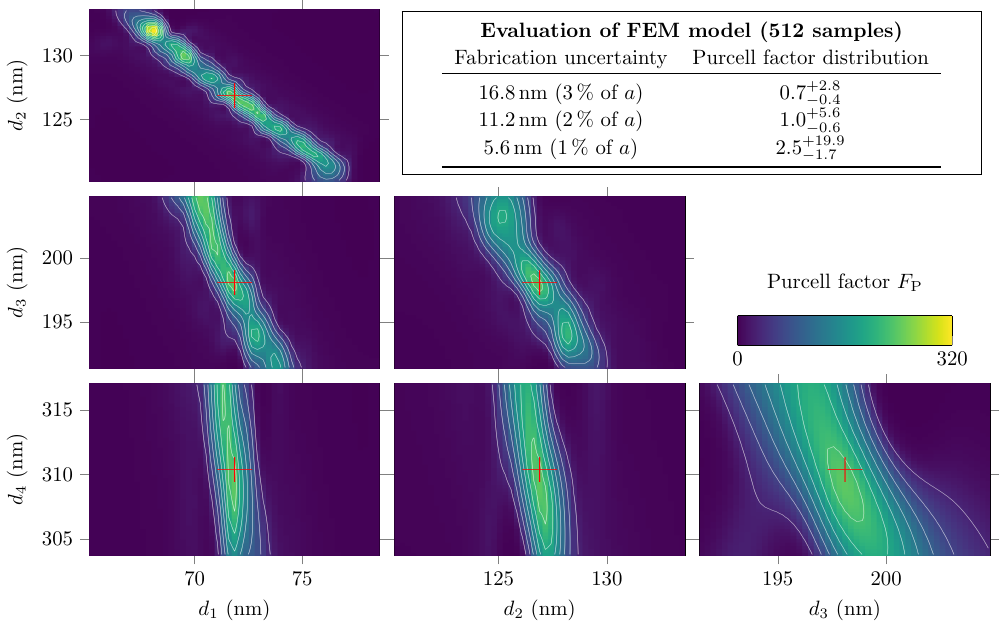}
  \caption{\label{fig:optima_comparison} Following the initial sampling and optimization
  of the complete parameter space (as given in \cref{tab:full_space}) an optimum
  $\vec{p}_{\mathrm{naive}} = \left[ 71.9, 126.9, 198.1, 310.4 \right]\,\si{\nano\meter}$
  was determined. This optimum is part of a very narrow ridge. When uncertainties in the
  manufacturing process are taken into consideration, the theoretically extremely high
  Purcell enhancements become difficult to achieve in reality. For example, assuming that
  the manufacturing process scatters according to a normal distribution with a standard
  deviation of \SI{16.8}{\nano\meter} in each parameter, one can realistically expect a
  median Purcell enhancement of $\overline{F}_{\mathrm{P}} = 0.7_{-0.4}^{+2.8}$, despite
  selecting a parameter that should theoretically reach values of $F_{\mathrm{P}} > 200$.}
\end{figure*}
\subsection{Comparing to the na\"ive ansatz}
\label{sec:comparing_to_naive}
The initial generation of training data $\vec{Y}_{\mathrm{train,first}}$ in the beginning
of the first pass can be considered to be part of a conventional optimization procedure
which seeks to maximize the Purcell enhancement of the photonic crystal nanobeam cavity in
the parameter space \cref{tab:full_space}. To complete this optimization, we took the
largest function value in $\vec{Y}_{\mathrm{train,first}}$ and the associated parameter,
and converged them using a Bayesian optimizer. The employed objective function was the FEM
model $f(\vec{p})$, as such no optimization w.r.t. robustness took place. The BO method
determined a parameter $\vec{p}_{\mathrm{naive}} = \left[ 71.9, 126.9, 198.1, 310.4
    \right]\,\si{\nano\meter}$ associated with a Purcell enhancement of $F_{\mathrm{P}}
  \approx \num{256.5}$.

We have calculated the expected Purcell enhancement of the found parameter
$\vec{p}_{\mathrm{naive}}$ by applying the methodology described in \cref{sec:mc}, where
$M_{\mathrm{device}} = \num{512}$ samples were used to evaluate the FEM model
$f(\vec{p})$. We found that given a manufacturing uncertainty of $\sigma_{d_{i}} =
  \SI{16.8}{\nano\meter}$ that even despite the very high Purcell enhancement of
$F_{\mathrm{P}} \approx \num{256.5}$, we can only expect to achieve a Purcell enhancement
of $\overline{F}_{\mathrm{P}} = 0.7_{-0.4}^{+2.8}$ when actually manufacturing the device.
This value can be improved by reducing the manufacturing uncertainties. For a
manufacturing uncertainty of \SI{11.2}{\nano\meter}, i.e., \SI{2}{\percent} of the unit
cell pitch $a$, we achieved $\overline{F}_{\mathrm{P}} = 1.0_{-0.6}^{+5.6}$, and for a
manufacturing uncertainty of \SI{5.6}{\nano\meter}, i.e., \SI{1}{\percent} of the unit
cell pitch $a$, we achieved $\overline{F}_{\mathrm{P}} = 2.5_{-1.7}^{+19.9}$.

We have visualized the close surroundings of the found optimum $\vec{p}_{\mathrm{naive}}$
in \cref{fig:optima_comparison}. Note that, in the creation of the figure, parameters with
even higher Purcell enhancements ($F_{\mathrm{P}} > \num{300}$) were discovered. This
highlights the challenges associated with optimizing the forward model function
$f(\vec{p})$ in such a large parameter space, even in an unrobust fashion. The size of the
shown parameter space is such that the $1\sigma$ ellipsis of a \SI{5.6}{\nano\meter}
manufacturing uncertainty would touch the borders of the sub figures. We observed that the
found maximum is very narrow. Manufacturing a cavity that exploits this maximum thus
requires very good control over the manufacturing uncertainties.

\section{Summary}
\label{sec:conclusion}

We have presented an approach for a robust design optimization that takes manufacturing
uncertainties in the fabrication process into account. We have applied it to optimize the
FEM model of a hole based photonic crystal nanobeam cavity such that we obtain a Purcell
enhancement that is robust under an assumed manufacturing uncertainty of $\sigma_{d_{i}} =
  \SI{16.8}{\nano\meter}$ for each of the varied device parameters.

The approach is based on a Monte Carlo integration procedure that is accelerated by using
trained Gaussian process surrogate models. These help to manage the high computational
burden imposed by the employed finite element forward model function $f(\vec{p})$. The GP
surrogates are adapted to deal with the bounded nature of the optimized Purcell
enhancement. During the application, a transformation function is learned that maps the
training data and the GPs predictions between the bounded domain of the training data and
the unbounded domain of the GP predictions.

The presented approach itself is iterative. First, a coarse analysis is performed on a
very large domain. The results of this coarse analysis are used to inform a finer analysis
that focuses on a much smaller domain. During the coarse analysis, we found that a region
centered around $\vec{p}_{\mathrm{opt,first}} = \left[ 428.6, 282.5, 369.5, 253
    \right]\,\si{\nano\meter}$ is predicted to have a robust Purcell enhancement of
$\overline{F}_{\mathrm{P}} \approx \num{9.4}$ (based on the coarse GP surrogate).
Evaluating the forward model function attested a robust Purcell enhancement of
$\overline{F}_{\mathrm{P}} \approx \num{2.3}$. Considering the very coarse nature of the
GP surrogate, we deemed this as a reasonable agreement, since large portions of the
parameter space have small~$F_{\mathrm{P}}$. During the fine analysis of the region
centered around $\vec{p}_{\mathrm{opt,first}}$, we found that the parameter
$\vec{p}_{\mathrm{robust}} = \left[ 412, 311.5, 392, 250.3 \right]\,\si{\nano\meter}$ is
robust against a manufacturing uncertainty of $\sigma_{d_{i}} = \SI{16.8}{\nano\meter}$,
and is capable of reaching a robust Purcell enhancement of $\overline{F}_{\mathrm{P}} =
  (3.6 \pm 1.1)_{-2.7}^{+6.6}$. The parameter was verified using the forward model function
$f(\vec{p})$. Here, we found that $\overline{F}_{\mathrm{P}} = 3.2_{-2.3}^{+8.6}$, which
confirmed the predictions of the GP surrogate within the prediction uncertainty and is an
improvement over the results of the first pass.

The result was compared to a na\"ive ansatz in which we directly optimized the forward
model function $f(\vec{p})$, i.e., found a candidate $\vec{p}_{\mathrm{naive}} = \arg\max
  f(\vec{p})$ and then determined the robustness of the found parameter. Here, we found
that, despite the fact that the $\vec{p}_{\mathrm{naive}}$ we found is associated with a
Purcell enhancement of $F_{\mathrm{P}} \approx \num{256.5}$, we can only expect a
\emph{robust} Purcell enhancement of $\overline{F}_{\mathrm{P}} = 0.7_{-0.4}^{+2.8}$ when
taking manufacturing uncertainties into account. The reason for this is that
$\vec{p}_{\mathrm{naive}}$ is situated in a very narrow ridge. In order to exploit the
high Purcell enhancement tied to this ridge, the manufacturing uncertainties have to be
controlled to a much tighter degree than assumed in this article.

We have therefore shown that our approach is capable of generating a result that is more
than four times better than the na\"ive approach when manufacturing uncertainties play a
role.

\section*{Acknowledgments}
We acknowledge funding by the German Federal Ministry of Education and Research (BMBF
project siMLopt number 05M20ZAA and BMBF Forschungscampus MODAL number 05M20ZBM) as well
as funding by the Deutsche Forschungsgemeinschaft (DFG, German Research Foundation) under
Germany's Excellence Strategy -- The Berlin Mathematics Research Center MATH+ (EXC-2046/1,
project ID: 390685689). This project (20FUN05 SEQUME) has received funding from the EMPIR
programme co-financed by the Participating States and from the European Union’s Horizon
2020 research and innovation programme. This project is co-financed by the European
Regional Development Fund (EFRD, application no. 10184206, QD-Sense).

\section*{Disclosures}

The authors declare no conflicts of interest.

\section*{Data availability}

The Python scripts for performing the robust design optimization, as well as the research
data, are available on Zenodo in the data publication found under \cite{zenodo_paper}.
Here we also publish additional information pertaining to the numerical model of the
photonic crystal nanobeam cavity, such as convergence plots or field exports.

\bibliographystyle{IEEEtran}
\bibliography{references}

\newpage
\appendix

\section*{Appendix}

\renewcommand{\thesubsection}{\Alph{subsection}}

\subsection{Variation of the fabrication uncertainties}
\label{secapp:variation}

\renewcommand\thefigure{\thesubsection.\arabic{figure}}
\setcounter{figure}{0}

\begin{figure*}[b]
  \includegraphics{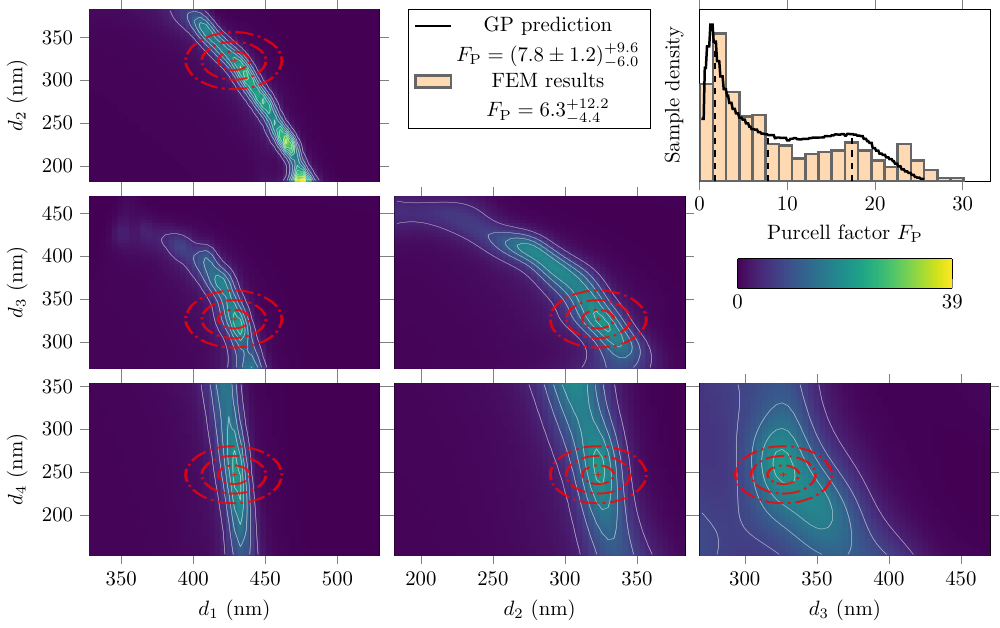}
  \caption{\label{fig:two_percent} The function value landscape around the robust design
  parameter $\vec{p}_{\mathrm{robust}}^{\prime} = \left[ 428.1, 323.9, 326, 246.8
      \right]\,\si{\nano\meter}$ (marked with a red cross). The red ellipses around the red
  cross indicate (from smallest to largest) the $1\sigma$, $2\sigma$, and $3\sigma$
  percentiles of the manufacturing distribution, which was assumed to be a multivariate
  normal distribution. The covariance matrix of this multivariate normal distribution was
  chosen to be $\mat{\Sigma}_{\mathrm{manuf}} = \diag(\sigma_{d}^{2}, \sigma_{d}^{2},
    \sigma_{d}^{2}, \sigma_{d}^{2})$ with $\sigma_{d} = \SI{11.2}{\nano\meter}$. Applying
  \cref{alg:mc_int} in conjunction with a trained surrogate model allowed us to
  efficiently determine that a manufacturing process that produces devices according to a
  normal distribution with $\mathcal{N}(\vec{p}_{\mathrm{robust}}^{\prime},
    \mat{\Sigma}_{\mathrm{manuf}})$ yields devices with an expected Purcell enhancement
  $\overline{F}_{\mathrm{{P}}} = (7.8 \pm 1.6)_{-6.0}^{+10.2}$. Verification using the FEM
  model showed that $\overline{F}_{\mathrm{P}}= 6.3_{-4.4}^{+12.2}$. Both results are
  shown in the top right sub figure.}
\end{figure*}
It is computationally relatively cheap to consider different manufacturing uncertainties
for an already trained surrogate model. We considered here the surrogate that was trained
in the second pass \cref{sec:results_second_pass}, which was defined on the domain listed
in \cref{tab:narrow_space}. If we assume that we can reduce the manufacturing uncertainty
to \SI{11.2}{\nano\meter} for each of the $d_{i}$ (i.e., to \SI{2}{\percent} of the unit
cell pitch $a$), then a parameter $\vec{p} = \left[ 428.1, 323.9, 326, 246.8
    \right]\,\si{\nano\meter}$ would lead to a predicted robust Purcell enhancement of
$\overline{F}_{\mathrm{P}} = (7.8 \pm 1.6)_{-6.0}^{+10.2}$. The results were verified
using FEM, with $\overline{F}_{\mathrm{P}} = 6.3_{-4.4}^{+12.2}$. In either case the
expected Purcell enhancement was approximately doubled. The results are shown in
\cref{fig:two_percent}.

\subsection{Bayesian optimization}
\label{sec:bo}

A very prominent application area for GPs is Bayesian optimization
(BO)~\cite{garnett_bayesoptbook_2023,schneider2019benchmarking,garcia2018shape}. BO
methods are a class of global sequential optimization methods that are known for being
very efficient at optimizing expensive black box functions (expensive in terms of the
consumed resources per evaluation)~\cite{jones1998efficient}. During the optimization, a
GP is iteratively trained on observations of the black box function $f(\vec{p})$. At each
iteration $M$ the predictions of the GP are used to generate a new sample candidate
$\vec{p}_{M+1}$, that is used to evaluate $f(\vec{p})$ again. These new results are used
to retrain the GP. New sample candidates are generated by maximizing a utility function
$\alpha(\vec{p})$ that assigns usefulness to points in the parameter space for achieving
the goal of the optimization, e.g., for maximizing or minimizing the optimized function.
Here, we employ the expected improvement (EI)~\cite{jones1998efficient} with respect to
the previously found smallest function value $f_{\mathrm{min}} = \min \{ f(\vec{p}_{1}),
  \dots, f(\vec{p}_{M})\}$, i.e., calculate
\begin{gather}
  \label{eq:utility}
  \vec{p}_{M+1} = \underset{\vec{p} \in \mathcal{X}}{\arg \max} \, \alpha_{\mathrm{EI}}(\vec{p}) \,, \\
  \text{with} \quad \alpha_{\mathrm{EI}}(\vec{p}) = \ex{ \min \left( 0,
    f_{\mathrm{min}} - \hat{f}(\vec{p}) \right)} \,.
\end{gather}
This iterative procedure is continued until a provided optimization is exhausted. The BO
method applied in this article is implemented in the analysis and optimziation
toolkit~\cite{schneider2019benchmarking} of the solver
JCMsuite~\cite{Pomplum_NanoopticFEM_2007}.

\end{document}